\documentclass[12pt,titlepage,final]{article}
\usepackage{axodraw,epsfig}
\newcommand{\tfrac}[2]{\textstyle\frac{#1}{#2}}
\newcommand{\half}{\tfrac12}

\begin{document}
  \begin{titlepage}
    \begin{flushright}
{\small\sf LPM/02-07\\
      UPRF-2002-02}
    \end{flushright}
\vskip 1.in
    \begin{center}
\textbf{\Large Boundary One-Point Functions, Scattering Theory and Vacuum 
Solutions in Integrable Systems}\\[2.em]
\textbf{\large V. A. Fateev\footnote{Laboratoire de Physique 
Math\'{e}matique, Universit\'{e} Montpellier II, Pl. E.Bataillon, 34095 
Montpellier, France.}$^,$\footnote{On leave of absence from Landau 
Institute for Theoretical Physics, ul.Kosygina 2, 117940 Moscow, Russia.}
\hspace{0.2cm} and E. Onofri\footnote{Dipartimento di Fisica, Universit\`a di 
Parma, and {\small\sf I.N.F.N.}, Gruppo Collegato di Parma, 43100 Parma, Italy}
}\\[5.em]

    \end{center}

\date{}

\bigskip
\bigskip
\begin{center}
\textbf{Abstract}
\end{center}
\bigskip
 Integrable boundary Toda theories are
considered. We use boundary one-point functions and boundary scattering
theory to construct the explicit solutions corresponding to classical vacuum
configurations. The boundary ground state energies are conjectured.
  \end{titlepage}

\section{ Introduction} \label{sec:introduction}

There is a large class of 2D quantum field theories (QFTs) which can be
considered as perturbed conformal field theories (CFTs). These theories are
completely defined if one specifies its CFT data and the relevant operator
which plays the role of perturbation. The CFT data contain explicit
information about the ultraviolet (UV) asymptotics of the field theory,
while the long distance behavior requires careful analysis. If a perturbed
CFT contains only massive particles it is equivalent to a relativistic
scattering theory and it is completely defined by specifying the S-matrix.
Contrary to CFT data, the S-matrix data exhibit information about the long
distance properties of the theory in an explicit way while the UV
asymptotics have to be derived.

A link between these two kinds of data would provide a good viewpoint for
understanding the general structure of 2D QFT. In general this problem does
not look tractable. Whereas the CFT data can be specified in a relatively
simple way the general S-matrix is a rather complicated object even in two
dimensions. However, there exists a rather important class of 2D QFTs
(integrable theories) where the scattering theory is factorized and the
S-matrix can be described in great detail. These theories admit a rather
complete description in the UV and IR regimes.

In this paper we consider the application of the boundary one-point functions
and boundary scattering theory to the explicit construction of the classical 
solutions describing the boundary vacuum configurations. As an example
of an integrable QFT we study the simply laced affine Toda theory (ATT),
which can be considered as a perturbed CFT (a non-affine Toda theory). This
CFT possesses an extended symmetry generated by a $W$-algebra. We consider 
the boundary conditions that preserve this symmetry. This permits us to apply
the ``reflection amplitude'' approach for the calculation of the 
boundary vacuum expectation values (VEVs) of exponential fields in ATTs.  
The reflection amplitudes in a CFT define the linear transformations
between different exponential fields corresponding to the same primary field
of the full symmetry algebra of theory. They play a crucial role in the
calculation of VEVs of the exponential fields in
perturbed CFTs \cite{FLZ} as well as for the analysis of UV asymptotics of
the observables in these QFTs \cite{ZZ},\cite{AFR}. The boundary VEVs, in 
particular, contain the information about the boundary values of the solutions 
of classical boundary ATTs equations. The information about the long distance 
behavior of these solutions can be extracted from the boundary $S$-matrix. 
The explicit solutions which are constructed in this paper provide 
us the consistency 
check of CFT and $S$-matrix data in integrable  boundary ATTs.

The plan of the paper is as follows: in section
\ref{sec:affine-non-affine} we recall some basic facts about Toda
theories and one-point functions in ATTs defined on the whole plane.
In section \ref{sec:bound-toda-theor} we consider integrable boundary
ATTs. The boundary one-point functions are used to derive the boundary
values of the solutions describing vacuum configurations in ATTs. We
calculate the classical boundary ground state energies and give the
conjecture for them in the quantum case. In section
\ref{sec:boundary-s-matrix} we describe the boundary scattering
theory, which is consistent with this conjecture, and construct the
boundary state.  The vacuum solutions describe the semiclassical
asymptotics of the one-point functions of the bulk operators in
boundary QFTs. This makes it possible to derive the long distance
asymptotics of vacuum solutions. These asymptotics determine
completely the explicit form of solutions which we construct in
section \ref{sec:boundary-solutions}. The solutions can be written in
terms of $\tau$-functions, associated with multisoliton solutions of
ATTs.  In the last section we consider the opposite (dual) limit of
the quantum correlation function of the bulk field with a boundary and
check that its boundary value coincides with that given by boundary
VEV. At the end of the section we consider the solutions for
non-simply laced ATTs which can be obtained by reduction from the
solutions for simply laced cases.

\bigskip

\section{ Affine and Non-Affine Toda Theories} \label{sec:affine-non-affine}

The ATT corresponding to a Lie algebra $G$ of rank $r$ is described by the
action:

\begin{equation}
\mathcal{A}=\int d^{2}x\left[ \frac{1}{8\pi }(\partial _{\mu }\varphi
)^{2}+\mu \sum_{i=1}^{r}e^{be_{i}\cdot \varphi }+
\mu e^{be_{0}\cdot \varphi }\right] ,  \label{at}
\end{equation}
where $e_{i},$ $i=1,...,r$ are the simple roots of $G$ and 
$-e_{0}$ is a maximal root:

\begin{equation}
e_0+\sum_{i=1}^rn_ie_i=0.  \label{mr}
\end{equation}
The fields $\varphi $ in Eq. (\ref{at}) are normalized so that at $\mu =0$

\begin{equation}
\left\langle \varphi _a(x)\,\varphi _b(y)\right\rangle =-\delta _{ab}\log
\left| x-y\right| ^2  \label{n}
\end{equation}
We will consider later the simply laced Lie algebras $A,D,E.$

For real $b$ the spectrum of these ATTs consists of $r$ particles which can
be associated with the nodes of the Dynkin diagram 
(or fundamental representations)
of $G$. The masses of these particles $m_{i}\ (i=1,...,r)$ are given by:

\begin{equation}
m_i=m\nu _i,\qquad m^2=\frac 1{2h}\sum_{i=1}^rm_i^2,  \label{s}
\end{equation}
where $h$ is the Coxeter number of $G$ and $\nu _i^2$ are the eigenvalues of
the mass matrix:

\begin{equation}
M_{ab}=\sum_{i=1}^rn_i(e_i)^a(e_i)^b+(e_0)^a(e_0)^b.  \label{mab}
\end{equation}

The exact relation between the parameter $m$ characterizing the spectrum of
physical particles and the parameter $\mu $ in the action (\ref{at}) can be
obtained by the Bethe ansatz method (see for example \cite{ALZ},\cite{F}).
It can be easily derived from the results of Ref. \cite{F} and has the form:

\begin{equation}
-\pi \mu \gamma (1+b^2)=\left[ \frac{mk(G)\Gamma \left( \frac {1-x}{h}
\right) \Gamma \left(\frac{x}{h}\right) x }{2\Gamma \left( \frac
1h \right)h }\right] ^{2(1+b^2)},  \label{mmu}
\end{equation}
where as usual $\gamma (z)=\Gamma (z)/\Gamma (1-z)$; $x=b^2/(1+b^2)$ and

\begin{equation}
k(G)=\left( \prod_{i=1}^rn_i^{n_i}\right) ^{1/2h}  \label{k}
\end{equation}
with $n_i$ defined by the equation (\ref{mr}).

The ATTs can be considered as perturbed CFTs. Without the last term (with
root $e_0$) the action (\ref{at}) describes NATTs, which are conformal. To
describe the generator of conformal symmetry we introduce the complex
coordinates $z=x_1+ix_2,\ \bar{z}=x_1-ix_2$ and the vector

\begin{equation}
Q=(b+1/b)\rho ,\qquad \rho =\half\sum_{\alpha >0}\alpha ,  \label{q}
\end{equation}
where the sum in the definition of the Weyl vector $\rho $ runs over all
positive roots $\alpha $ of Lie algebra $G$.

The holomorphic stress energy tensor

\begin{equation}
T(z)=-\half(\partial _{z}\varphi )^{2}+Q\cdot \partial _{z}^{2}\varphi
\label{se}
\end{equation}
ensures the local conformal invariance of the NATT.

Besides conformal invariance NATT possesses an additional symmetry generated
by two copies of the chiral $W(G)$-algebras: $W(G)\otimes $ $\overline{W}(G)$. 
The full chiral $W(G)$-algebra contains $r$ holomorphic fields $W_{j}(z)$ 
$(W_{2}(z)=T(z))$ with spins $j$ which follow the exponents of the Lie
algebra $G.$ The explicit representation of these fields in terms of fields
 $\partial _{z}\varphi _{a}$ can be found in \cite{FL}.
The exponential fields:

\begin{equation}
V_{a}(x)=e^{a\cdot \varphi (x)}  \label{ex}
\end{equation}
are spinless conformal primary fields with dimensions $\Delta (a)=
\frac{Q^{2}-(a-Q)^{2}}{2}$. The fields $V_{a}(x)$ are also primary 
fields with respect
to full chiral algebra $W(G)$. It was shown in Ref. \cite{FL} that fields
 $V_{a}(x)$ and $V_{s(a)}(x)$ correspond to the same representation of 
$W(G)$-algebra if their parameters are related by any transformation of the 
Weyl group $\mathcal{W} $ of the Lie algebra $G$ acting on the vector $a$ as 
follows:

\begin{equation}
a\rightarrow s(a)=Q+\mathbf{\hat{s}}(a-Q)\mathbf{\ ,\quad \hat{s}\in }
\mathcal{W}\,.  \label{wg}
\end{equation}
This means that the fields $V_{s(a)}$ for different
$\mathbf{\hat{s}}\in \mathcal{W}$ are the reflection image of each
other and are related by a linear transformation.  The functions that
define this transformation are known as ``reflection
amplitudes''. Reflection amplitudes play an important role in the
analysis of perturbed CFTs. In particular they permit to calculate the
vacuum expectation values of exponential fields in ATTs:

\begin{equation}
G(a)=\left\langle \exp a\cdot \varphi \right\rangle  \label{g}
\end{equation}
These\ VEVs were calculated in Ref.\cite{VFA} and they can be represented 
in the form:

\begin{equation}
G(a) =\left[ \frac{mk\Gamma \left( \frac{1-x}{h}\right) \Gamma
\left( \frac{x}{h}\right) x}{2\Gamma \left( \frac{1}{h}
\right) h}\right] ^{-a^{2}}
\exp \left( \int\limits_{0}^{\infty }\frac{dt}{t}
[a^{2}e^{-2t}-F(a,t)]\right)  \label{vev}
\end{equation}
with

\begin{eqnarray}
F(a,t) &=&\sinh ((1+b^{2})t)  \nonumber \\
&&\times \sum_{\alpha >0}\frac{\sinh (ba_{\alpha }t)\sinh ((b(a-2Q)_{\alpha
}+h(1+b^{2}))t)}{\sinh t\sinh (b^{2}t)\sinh ((1+b^{2})ht)}  \label{fb}
\end{eqnarray}
here and below subscript $\alpha $ denotes the scalar product of the vector
with a positive root $\alpha ,$ i.e.:

\begin{equation}
a_{\alpha }=a\cdot \alpha ;\quad (a-2Q)_{\alpha }=(a-2Q)\cdot \alpha .
\label{aalf}
\end{equation}

This expression satisfies many possible perturbative and non-perturbative
tests for one-point function in ATT. For example it can be easily derived
from Eq.(\ref{mmu}) and from the equation of motion that the 
bulk vacuum energy $E(b)$ in
ATT is expressed in terms of function $G(a)$ as:

\begin{equation}
n_i E(b)=h(1+b^2)\mu G(be_i).  \label{ebulk}
\end{equation}
The values of the function $G(a)$ at the special 
points $be_i$ can be calculated
explicitly and the result coincides with the known expression for the bulk
vacuum energy \cite{DdV}:

\begin{equation}
E(b)=\frac{m^{2}\sin (\pi /h)}{4\sin (\pi x/h)\sin (\pi (1-x)/h)}
\label{ebul}
\end{equation}
where $x=b^{2}/(1+b^{2})$.

It is easy to see from Eq.(\ref{ebulk}) that function $E(b)$ (and the 
function $G(a)$ as well) is invariant under duality 
transformation $b\rightarrow 1/b$
or $x\rightarrow 1-x$. The same is true for the bulk scattering $S$-matrix
in ATT. This remarkable property of duality is very important for the
analysis of Toda theories.

In the limit $b\rightarrow 0$ the field $\widetilde{\varphi }=b\varphi $ can
be described by the classical equations. In particular it can be derived
from Eq.(\ref{vev}) that the VEV of this field 
$\widetilde{\varphi }_{0}=\left\langle b\varphi \right\rangle $ 
at that limit can be expressed in
terms of the fundamental weights $\omega _{i}$ of $G$ as:

\begin{equation}
\tilde{\varphi}_{0}=b\varphi _{0}=\sum_{i=1}^{r}(\log n_{i}-2\log
k(G))\omega _{i}  \label{clv}
\end{equation}
and coincides with a classical vacuum of ATT (\ref{at}). At this limit we
have (see Eq.(\ref{mmu})) that $\mu =(mk(G))^{2}/4\pi b^{2}+O(1)$ and after
rescaling and shifting \ of the field $\varphi $ the action (\ref{at}) can
be expressed in terms of the field $\phi =b\varphi -\tilde{\varphi}_{0}$ as:

\begin{equation}
\mathcal{A}_{b}=\frac{1}{4\pi b^{2}}\int d^{2}x\left[ \half(\partial
_{\mu }\phi )^{2}+m^{2}\sum_{i=0}^{r}n_{i}e^{e_{i}\cdot \phi }\right] +O(1).
\label{aTTT}
\end{equation}
The VEVs given by Eq.(\ref{vev}) and scattering theory data were used
in Ref. \cite{VFA} to describe the asymptotics of the cylindrical
solutions of classical Toda equations following from the action given
by Eq.(\ref{aTTT}). In the next sections we use a similar strategy to
explicitly find the vacuum solutions in integrable boundary ATT.

\section{Boundary Toda Theories, One-Point Functions and Classical Vacuum
Solutions} \label{sec:bound-toda-theor}

In this section we consider simply laced boundary Toda theories defined on the
half-plane $H=(x,y;y>0)$.
The integrability conditions for classical simply laced ATT on $H$ were
studied in the paper\cite{BCD}. It was shown there that the action of
integrable boundary ATT can be written as:

\begin{eqnarray}
\mathcal{A}_{bound}&=&\int\limits_Hd^2x\left[ \frac 1{8\pi }(\partial _\mu
\varphi )^2+\mu \sum_{i=1}^re^{be_i\cdot \varphi }+
\mu e^{be_0\cdot \varphi }\right]  \nonumber \\
&& +\mu _B\int dx\sum\limits_0^r s_i e^{be_i\cdot\varphi /2}  \label{atb}
\end{eqnarray}
where either all the parameters $s_i$ vanish, corresponding to the Neumann
boundary conditions:

\begin{equation}
\partial _y\varphi (x,0)=0;  \label{N}
\end{equation}
or they are given by $s_i=\pm 1$ and the parameter $\mu _B$ is related to the
parameter $\mu $ in the bulk (in the classical case) as:

\begin{equation}
\mu _B^2=\mu /\pi b^2.  \label{ND}
\end{equation}
For the Lie algebra $A_1$ (sinh-Gordon model) the integrability conditions are
much less stringent and the parameters $s_0$ and $s_1$ can have arbitrary
values \cite{GZ}. 
Here we discuss this problem for other Lie algebras where the choice of
integrable conditions is rather restrictive. We consider Toda theories with
Neumann boundary conditions and in the case when all parameters $s_i=1$
(with $s_0=0$ in the non-affine case). Really, these two quite different
classical theories in the quantum case are described by the same theory and
are related by a duality transformation ($b\rightarrow 1/b$) \cite{CRG}, 
\cite{GHB}. The cases corresponding to different signs of the parameters 
$s_i$ are more subtle and will be considered elsewhere.

We start from a consideration of the boundary NATTs described by the action 
(\ref{atb}) without the last term in the bulk action and with parameter 
$s_{0}=0$ in the boundary term. The boundary ATTs will be considered as
perturbed boundary CFTs. On the half-plane with $W$-invariant boundary
conditions we have only one $W$-algebra. In this case the currents 
$\overline{W}_{j}(\overline{z})$ should be the 
analytical continuations of
the currents $W_{j}(z)$ into the lower half-plane. In particular, they
should coincide at the boundary. These conditions impose very strong
restrictions on the form of the boundary terms in the action. It is rather
easy to derive from the explicit form of $W$-currents \cite{FL} that the
Neumann boundary conditions (\ref{N}) preserve $W$-symmetry. The boundary
condition (\ref{ND}) whose quantum modified version has the form \cite{FZZA}

\begin{equation}
\mu _{B}^{2}=\frac{\mu }{2}\cot \left( \frac{\pi b^{2}}{2}\right) ,
\label{MND}
\end{equation}
(with all $s_{i}=1,i=1,...r$ and $s_{0}=0$) describes the dual theory and,
hence, also preserves $W-$symmetry.

In the $W$-invariant boundary NATT we have two kinds of exponential fields.
The bulk fields $V_{a}(x,y)$ and the boundary fields $B_{a}(x)$ are defined
as:

\begin{equation}
V_{a}(x,y)=\exp \{a\cdot \varphi (x,y)\};\quad B_{a}(x)=\exp \{\half a\cdot
\varphi (x)\}.  \label{VA}
\end{equation}
These fields are specified by the same quantum numbers as those for the
corresponding fields (\ref{ex}) defined on the whole plane. In particular,
they possess the similar properties with respect to the action of the Weyl
group of $G$ (see section 2), defined by Eq.(\ref{wg}). Following the
standard lines one can introduce and calculate boundary reflection
amplitudes which relate the fields $B_{a}(x)$ and $B_{s(a)}(x)$ 
(see Ref.\cite{VALF} for details).

Boundary reflection amplitudes can be used for the calculation of vacuum
expectation values of the boundary exponential fields in ATTs. One-point
functions:

\begin{equation}
G_{B}(a)=\left\langle \exp (\half a\cdot \varphi)\right\rangle _{B}  \label{OP}
\end{equation}
for $ADE$ ATTs were calculated in \cite{VLF} and have the form:

\begin{equation}
G_{B}(a) =\left[\frac{mk\Gamma \left(\frac{1-x}{h}\right) \Gamma
\left( \frac{x}{h}\right) x}{2\Gamma \left( \frac{1}{h}\right) h}\right]
^{-a^{2}/2}
\exp \left( \int\limits_{0}^{\infty }\frac{dt}{t}[\frac{a^{2}}{2}
e^{-2t}-F_{B}(a,t)]\right)  \label{vevs}
\end{equation}
with

\begin{equation}
F_B(a,t)=f(t)\sum_{\alpha >0}\frac{\sinh (ba_\alpha t)\sinh ((b(a-2Q)_\alpha
+h(1+b^2))t)}{\sinh 2t\sinh (2b^2t)\sinh ((1+b^2)ht)}  \label{fbs}
\end{equation}
where for the boundary conditions (\ref{MND}) the function $f(t)$ is:

\begin{equation}
f(t)=2e^t\sinh ((1+b^2)t)\cosh (b^2t)  \label{fB}
\end{equation}
and for the dual theory which corresponds to 
Neumann boundary conditions (\ref{N}) we should make 
the substitution $f(t)\rightarrow f^{\vee}(t)$:

\begin{equation}
f^{\vee}(t)=2e^{tb^2}\sinh ((1+b^2)t)\cosh t.  \label{fN}
\end{equation}

It is easy to see from the explicit form of $G_{B}(a)$ that in the classical
limit ($b\rightarrow 0$ with $b\varphi $ is fixed) the boundary VEV 
$\widetilde{\varphi }_{0,B}$ of the field $b\varphi $ for the Neumann
boundary conditions coincides with the classical 
vacuum $\widetilde{\varphi }_{0}$ (\ref{clv}) in the bulk. 
For the boundary conditions (\ref{MND}) this
happens only for Lie algebra $A_{r}$, where both these values vanish. For
other cases we can derive from Eqs.(\ref{vevs}-\ref{fB}) that difference of
VEVs of the field $b\varphi $ at the boundary and on the whole plane has a
form:

\begin{equation}
\Theta =\left\langle b\varphi \right\rangle _{B}-\left\langle b\varphi
\right\rangle =-\sum\limits_{\alpha >0}\alpha \int\limits_{0}^{\infty }dt
\frac{x\sinh ((h-2\rho _{\alpha })t)\sinh t}{\sinh xt\cosh (1-x)t\sinh ht}
\label{TET}
\end{equation}
Vector $\Theta $ can be expressed in terms of elementary functions at the
semiclassical limit $x\rightarrow 0$, at the self-dual point $x=1/2$ and in
the dual limit $x\rightarrow 1$. We discuss the dual limit in the last
section. Here we note that at $x\rightarrow 0$ vector $\Theta $ has a limit:

\begin{equation}
\vartheta =-\sum\limits_{\alpha >0}\alpha \int\limits_{0}^{\infty }
\frac{dt}{t}\frac{\sinh ((h-2\rho _{\alpha })t)}
{\sinh (ht)}\tanh t.  \label{TT}
\end{equation}
For $D$ and $E$ series, which we consider below the numbers $2l_{\alpha
}=h-2\rho _{\alpha }$ are always even. For the following it is convenient to
define:

\begin{equation}
E_{i}=\exp (\half \vartheta \cdot e_{i})=\prod_{\alpha >0}(p_{l_{\alpha
}})^{-e_{i}\cdot \alpha }  \label{Ei}
\end{equation}
where $p_{l}=1/p_{-l}$ can be written in terms of trigonometric functions as:

\begin{equation}
p_{l}^{2}=\exp (\int\limits_{0}^{\infty }\frac{dt}{t}\frac{\sinh 2lt}{\sinh
ht}\tanh t)=\prod_{k=1}^{l}\frac{\cos (\pi (l+1-2k)/h)}{\cos (\pi (l-2k)/h)}.
\label{pl}
\end{equation}
In particular:

\begin{equation}
E_{0}=\frac{p_{q-1}p_{h/2-q}}{p_{h/2-1}}  \label{E0}
\end{equation}
here $q(G)=\max_{i}n_{i};\ q(D)=2,q(E_{6})=3,q(E_{7})=4,q(E_{8})=6$.

The vector $\vartheta $ is simply related to the boundary soliton solution that
describes the classical vacuum configuration. The classical problem for this
solution $\phi (y),y>0$ can be formulated in the following way. We are
looking for a solution to the classical Toda equations, which decreases at 
$y\rightarrow \infty $ and satisfies at $y=0$ the boundary conditions that
follow from the action (\ref{atb}). After rescaling and shifting (see
section 2) the field $\phi =b\varphi -$ $\widetilde{\varphi }_{0}$ satisfies
the equation:

\begin{equation}
\partial _y^2\phi =m^2\sum_{i=0}^rn_ie_i\exp (e_i\cdot \phi );\qquad n_0=1
\label{EQT}
\end{equation}
and the boundary condition at $y=0$:

\begin{equation}
\partial _{y}\phi =m\sum\limits_{i=0}^{r}\sqrt{n_{i}}e_{i}\exp (\half
e_{i}\cdot \phi ).  \label{YN}
\end{equation}
The vector $\vartheta $ is equal to the boundary value of this solution and
hence, it completely fixes the boundary soliton. At $y=0$ we have

\begin{equation}
\phi (0)=\vartheta ;\quad \partial _{y}\phi (0)=\sum\limits_{i=0}^{r}
\sqrt{n_{i}}e_{i}E_{i}.  \label{bE}
\end{equation}
This gives us the possibility to study Eq.(\ref{EQT}) numerically. The
numerical analysis of this equation shows that for these and only for these
boundary values a smooth solution decreasing at infinity exist (see appendix
C).

It is natural to expect that solution $\phi (y)$ to the Eqs. (\ref{EQT}),
(\ref{YN}) can be expressed in terms of tau-functions associated with
multi-soliton solutions of the classical ATT equations (see, for example 
\cite{BOWC},\cite{BOW} and references there). 
For the $D$ and $E$ series of algebras (besides the
cases $D_{4}$ and $D_{5}$ \cite{BOWC} ) the explicit form of these solutions
was not known. We postpone the discussion of these 
solutions to section 5.
Here we consider the classical boundary ground state energy which can be
defined as:

\begin{equation}
\mathcal{E}_{bound}^{(cl)}=\frac{1}{4\pi b^{2}}[2m\sum_{i=0}^{r}\sqrt{n_{i}}
E_{i}+\int\limits_{0}^{\infty }dy(\half(\partial _{y}\phi
)^{2}+m^{2}\sum_{i=0}^{r}n_{i}(e^{e_{i}\cdot \phi }-1))].  \label{eBon}
\end{equation}
We note that numerical values of $\vartheta ^{2}$, defined by Eq.(\ref{TT})
are rather small for all $G$, and the functional (\ref{eBon}) can be
calculated to a good accuracy using the bilinear approximation:
\[
\mathcal{E}_{bound}^{(cl)}=\frac{m}{4\pi b^{2}}\sum_{i=0}^{r}\sqrt{n_{i}}
E_{i}(2-\log E_{i})+O(|\vartheta |^{3}). 
\]

More careful analysis of Eqs.(\ref{EQT}),(\ref{YN}) (see section 5) gives us
the reasons to write the expression for boundary ground state energy in the
following form. Namely, we denote as $\Sigma _{m}(G)$ the sum of the masses
of all particles in the ATT:

\begin{equation}
\Sigma _m(G)=\sum_{i=1}^rm_i=m\sum_{i=1}^r\nu _i=m\cdot tr(M^{1/2}).
\label{sIg}
\end{equation}
This has the following values for simply laced Lie algebras:

\begin{eqnarray}
\Sigma _m(A_{n}) &=&2m\cot (\pi /2h);\quad \Sigma _m(D_n)=\frac{2m\cos (\pi
/4-\pi /2h)}{\sin (\pi /2h)};  \nonumber \\
\Sigma _m(E_6) &=&m(6-2\sqrt{3})^{1/2} \frac{\cos (\pi /8)}{\sin (\pi /2h)};
\quad \Sigma _m(E_7)=\frac{2m\sin (2\pi /9)}{\sin (\pi /2h)};  \nonumber \\
\Sigma _m(E_8) &=&4m(\sqrt{3}\sin (\pi /5)\sin (\pi /30))^{1/2}\frac{\cos
(\pi /5)}{\sin \left( \pi /2h\right) }.  \label{sUm}
\end{eqnarray}
The classical boundary ground state energy (\ref{eBon}) can be written in
terms of these values as:

\begin{equation}
\mathcal{E}_{bound}^{(cl)}(G)=\frac h{4\pi b^2}\tan (\pi /2h)\Sigma _m(G).
\label{eCl}
\end{equation}

In the quantum case the boundary ground state energy 
$\mathcal{E}_{bound}^{(q)}$ will have contributions coming 
from the boundary term in the
Hamiltonian and from the bulk fluctuations around the background solution.
The contributions of the first type can be calculated using the explicit
expression for vacuum expectation values (\ref{vevs}). For small $b $ the
first quantum correction of the second type can be expressed through the
boundary $S-$matrix at $b=0$ (see, for example \cite{COD},\cite{COT}). We
will discuss these boundary $S$-matrices below. Here we note that in the
strong coupling region $b\gg1$ our theory is described by the weakly coupled
dual ATT with Neumann boundary conditions (\ref{N}). In the strong coupling
limit the dual theory is a set of $r$ free 
bosonic theories with masses $m_i$.  
The boundary ground state energy for the free massive bosonic theory with
Neumann boundary conditions and mass $m_i$ can be easily calculated and is
equal to $m_i/8$. The first perturbative correction in the weakly coupled
dual theory can be also evaluated with the result:

\begin{equation}
\mathcal{E}_{bound}^{(q)}(G)=\frac{\Sigma _m(G)}8\left( 1+\frac \pi
{2hb^2}\cot (\pi /2h)+O(1/b^4)\right) .  \label{eDu}
\end{equation}
Both asymptotics $b\rightarrow 0$ (\ref{eCl}) and $b\rightarrow \infty $ 
(\ref{eDu}) are in agreement with the following conjecture for boundary
ground state energy:

\begin{equation}
\mathcal{E}_{bound}^{(q)}(G)=\frac{\sin (\pi /2h)\Sigma _{m}(G)}{8\sin (\pi
x/2h)\cos (\pi (1-x)/2h)}  \label{eQ}
\end{equation}
where $x=b^{2}/(1+b^{2})$.

The nonperturbative check of this conjecture can be made using the boundary
Thermodynamic Bethe Ansatz equations \cite{SSM}. The kernels in these
nonlinear integral equations depend on the bulk and boundary $S$-matrices.
In particular boundary ground state energy can be expressed in terms of the
mass $m_{j}$ of the particle $j$ multiplied by the ratio of Fourier
transforms of logarithms of boundary $S$-matrix $R_{j}$ and bulk amplitude $%
S_{jj}$ taken at $\omega =i$. To check our conjecture and to construct the
explicit solution of Eq.(\ref{EQT}) we need the information about
boundary $S-$matrix.

\section{ Boundary S-matrix and Boundary State} \label{sec:boundary-s-matrix}

The boundary $S$-matrix (reflection coefficient) in the ATT for the particle 
$j$ corresponding to the fundamental representation $\pi _{j}(G)$ can be
defined as:

\begin{equation}
|j,-\theta \rangle _{out}=R_{j}(\theta )|j,\theta \rangle _{in}  \label{kI}
\end{equation}
where $\theta $ is the rapidity of particle $j$.

The reflection coefficients for the $A_{r}$ ATT with boundary conditions ( 
\ref{MND}) were conjectured in \cite{CODS},\cite{DELG}. They can be written
in terms of function:

\begin{equation}
\left( z\right) =\frac{\sin \left( \frac \theta {2i}+\frac \pi {2h}z\right)}
{\sin \left( \frac \theta {2i}-\frac \pi {2h}z\right) }  \label{Z}
\end{equation}
in the following way:

\begin{equation}
R_{j}(\theta )=\prod\limits_{a=1}^{j}(a-1)(a-h)(-a+x)(-a+h+1-x).  \label{aNk}
\end{equation}
Unfortunately, we were not able to find in the literature any conjecture for
other Lie algebras consistent with duality properties discussed above. So we
will give here a conjecture which naturally generalizes the reflection
coefficients (\ref{aNk}) to other simply laced Lie algebras. To do this we
rewrite Eq. (\ref{aNk}) in the form:

\begin{equation}
R_{j}(\theta )=\exp (-i\delta _{j}(\theta ))  \label{kJ}
\end{equation}
where

\begin{equation}
\delta _{j}=\int\limits_{0}^{\infty }
\frac{dt}{t}\sinh (\frac{2h\theta t}{\pi })
[\sinh ((1-x)t)\sinh ((h+x)t)\Delta _{j}(A_{r},t)-2]  \label{dJ}
\end{equation}
and

\begin{equation}
\Delta _{j}(A_{r},t)=\frac{8\sinh jt\sinh (h-j)t}{\sinh t\sinh 2ht}.
\label{DJ}
\end{equation}
The natural generalization of these equations can be written as:

\begin{equation}
R_{j}(\theta )=\Phi _{j}(\theta )\exp (-i\delta _{j}(\theta ))  \label{gNr}
\end{equation}
where $\Phi _{j}(\theta )$ are the CDD factors, satisfying the conditions:

\begin{equation}
\Phi _{j}(\theta )\Phi _{j}(-\theta )=1;\quad \Phi (\theta )\Phi (\theta
+i\pi )=1  \label{cDd}
\end{equation}
and the function $\delta _{j}\left( \theta \right) $ is defined by Eq.(\ref
{dJ}), where we should do the substitution $\Delta _{j}(A_{r},t)\rightarrow $
$\Delta _{j}(G,t)$ with:

\begin{equation}
\Delta _{j}(G,t)=\frac{4}{\cosh ht}\left[\left( (2\cosh t-2)\delta
_{mn}+e_{m}\cdot e_{n}\right) ^{-1}\right]_{jj}.  \label{cOn}
\end{equation}
The most important part of this conjecture is that for the particles $j$
corresponding to fundamental representations $\pi _{j}(G)$ (or the node $j$
of Dynkin diagram) with $n_{j}$ in Eq.(\ref{mr}) equal to $1$, the CDD
factors $\Phi _{j}(\theta )$ in Eq.(\ref{gNr}) are equal to one.

This statement fixes completely the boundary $S$-matrix for Lie algebras 
$D_{r},E_{6},E_{7}$. We denote by $R_{f}(G,\theta )$ the fundamental
reflection coefficients. This means that all other amplitudes $R_{j}(\theta
) $ can be obtained from the fundamental reflection factors $R_{f}(G,\theta
) $ by application of the boundary bootstrap fusion procedure \cite{GZ},\cite
{FK}. It is easy to see that $R_{f}(D_{r},\theta )=R_{r}(\theta
)=R_{r-1}(\theta )$, where particles $r$ and $r-1$ correspond to the spinor
representations of $D_{r}$; $R_{f}(E_{6},\theta )=R_{1}\left( \theta \right)
=R_{\overline{1}}\left( \theta \right) $, where particles $1$ and 
$\overline{1}$ form the doublet of lightest particles in the $E_{6}$ ATT and 
$R_{f}(E_{7},\theta )=R_{1}(\theta )$, where $1$ is the lightest particle in
the $E_{7}$ ATT. For all these three cases the CDD factors $\Phi _{f}(\theta
)=1$ and the reflection coefficients are given by Eqs.(\ref{kJ}),(\ref{dJ})
with $j=f$ and

\begin{equation}
\Delta _f(G,t)=\frac{8\sinh \left( ht/2\right) \sinh ((h/2+q-1)t)}{\sinh
(qt)\sinh (2ht)}  \label{dE}
\end{equation}
where $q(G)=\max_in_i$; $q(D)=2,q(E_6)=3,q(E_7)=4$.

As an example of the application of the boundary bootstrap equations we give
here the reflection coefficients $R_{j}$ for the particles $j=1,2,...,n-2$
in the $D_{r}$ ATT, which can be obtained from the amplitude $R_{f}\left(
\theta \right) =R_{r}(\theta )=R_{r-1}(\theta )$. These functions $R_j$
have the form (\ref{gNr}), where:

\begin{equation}
\Delta _{j}(D_{r},t)=\frac{16\sinh jt\cosh ((h/2-j)t)\sinh (ht/2)}{\sinh
t\sinh 2ht}  \label{D}
\end{equation}
and the CDD factor can be written as:

\begin{equation}
\Phi _{j}=\exp \left( i\int\limits_{0}^{\infty }\frac{dt}{t}
\sinh (2h\theta t/\pi )
\sinh (1-x)t\cosh (xt)\psi _{j}(D_{r},t)\right)  \label{D1}
\end{equation}
where $\psi _{j}(D_{r},t)=8\sinh (j-1)t\sinh jt/(\sinh 2t\cosh ht)$

The Lie algebra $E_{8}$ has no fundamental representations with $n_{j}=1$.
The lightest particle in this case is associated with the adjoint
representation and $R_{f}(E_{8},\theta )=R_{ad}(\theta )$. The adjoint
representations for Lie algebras $D,E$ have $n_{ad}=2$ and, hence, CDD
factors should appear. The reflection coefficients $R_{ad}(\theta )$ for Lie
algebras $D$ and $E$ can be written in the form (\ref{gNr}), with:

\begin{equation}
\Delta _{ad}(G,t)=\frac{8\cosh ((q-1)t)\cosh ((h/2-q)t)}
{\cosh (ht/2)\cosh ht}  \label{adj}
\end{equation}
where $q(G)$ is defined above and the CDD factor $\Phi _{ad}(\theta )$ is
defined by Eq.(\ref{D1}) with function

\begin{equation}
\psi _{ad}(G,t)=\frac{8\sinh t}{\cosh ht}  \label{ad1}
\end{equation}

The amplitudes $R_{j}(\theta )$ following from the scattering theory
described above have no poles corresponding to the bound states of particles 
$j$ with boundary. The only poles with positive residues which appear in
some of the functions $R_{j}(\theta )$ are the poles at $\theta =i\pi /2$

\begin{equation}
R_{j}(\theta )=\frac{iD_{j}^{2}(b)}{\theta -i\pi /2}  \label{p}
\end{equation}
corresponding to the particle boundary coupling with zero binding energy.
For example, the lightest particle whose boundary amplitude possess such
pole is the particle $m_{ad}$ corresponding to adjoint representation. It is
easy derive from Eqs.(\ref{adj},\ref{ad1}) that corresponding residue 
has a form:

\begin{equation}
D_{ad}^{2}(b)=2\cot \left( \frac{\pi x}{2h}\right) \tan \left( \frac{\pi
\left( 1-x\right) }{2h}\right) \tan \left( \frac{\pi }{2h}\right) 
\mathcal{T}_{0}\mathcal{T}_{q-1}\mathcal{T}_{h/2-q}  \label{res}
\end{equation}
where $\mathcal{T}_{k}=\tan (\frac{\pi }{4}-\frac{\pi (k+1)}{2h})
\tan (\frac{\pi }{4}-\frac{\pi (k+1-x)}{2h})
\cot \left( \frac{\pi }{4}-\frac{\pi k}{2h}
\right) \cot \left( \frac{\pi }{4}-\frac{\pi (k+x)}{2h}\right) $.

The amplitudes $R_{j}(\theta )$ and their residues can be used for the
construction of the boundary state in integrable QFT. It was shown in Ref. 
\cite{GZ} that this state can be written in terms of numbers $D_{j}$ and
functions $K_{j}(\theta )=R_{j}(i\pi /2-\theta )$ as:

\begin{equation}
|B\rangle =\exp \left[ \sum_{j=1}^{r}D_{j}(b)A_{j}(0)+\frac{1}{4\pi }\int
d\theta K_{j}(\theta )A_{j}(\theta )A_{j}(-\theta )\right] |0\rangle
\label{BS}
\end{equation}
where $A_{j}(\theta )$ are the operators creating the asymptotic states
with rapidity $\theta$ i.e. $A_{j}(\theta )|0\rangle =|j,\theta \rangle _{in}$.

We will be mostly interested by the linear in the operators $A_{j}$ part in
the expansion of boundary state $|B\rangle $
\[
|B\rangle =[1+\sum_{j=1}^{r}D_{j}(b)A_{j}(0)]|0\rangle +... 
\]
which determines ``one particle'' contributions to the correlation function:

\begin{equation}
\mathcal{F}_{b}\left( y\right) =\langle 0|b\varphi (y)|B\rangle -\langle
0|b\varphi |0\rangle  \label{Fg}
\end{equation}
where the variable $y>0$ plays the role of Euclidean time. These ``one
particle'' contributions can be written as:

\begin{equation}
\mathcal{F}_{b}\left( y\right) =\sum_{j=1}^{r}D_{j}(b)\langle 0|b\varphi
(0)|j,0\rangle _{in}\exp (-m_{j}y)+...  \label{opc}
\end{equation}
The boundary soliton $\phi (y)$ (see section 3) coincides with the
semiclassical limit of this correlation function i.e. $\phi
(y)=\lim_{b\rightarrow 0}\mathcal{F}_{b}\left( y\right) $. In the weak
coupling limit one particle matrix elements can be easily calculated and
have a form:

\[
\langle 0|\varphi (0)|j,0\rangle _{in}=\sqrt{\pi }\xi _{j}+O(b^{2}) 
\]
where $\xi _{j}$ are the eigenvectors of mass matrix: $M\xi _{j}=\nu
_{j}^{2}\xi _{j}$, satisfying the conditions:

\begin{equation}
\xi _{i}\cdot \xi _{j}=\delta _{ij};\quad \rho \cdot \xi _{j}\geq 0.
\label{eigv}
\end{equation}

The corresponding ``one particle'' contributions to the boundary solution 
$\phi (y)$ will be:

\begin{equation}
\phi (y)=\sum_{j=1}^{r}d_{j}\xi _{j}\exp (-m_{j}y)+...  \label{fld}
\end{equation}
where

\begin{equation}
d_{j}=\lim_{b\rightarrow 0}\sqrt{\pi }bD_{j}(b).  \label{dj}
\end{equation}
For example the coefficient $d_{ad}$ corresponding to the lightest particle 
$m_{ad}$ in expansion (\ref{fld}) and defining the main term of the
asymptotics at $y\rightarrow \infty $ can be extracted from Eq.(\ref{res})
and has a form:

\begin{equation}
d_{ad}=2\sqrt{h}\tan \left( \frac{\pi }{2h}\right) T_{0}T_{q-1}T_{h/2-q}
\label{dad}
\end{equation}
where $T_{k}^{2}$ $=\lim_{b\rightarrow 0}\mathcal{T}_{k}:$

\begin{equation}
T_{k}=\tan (\frac{\pi }{4}-\frac{\pi (k+1)}{2h})\cot \left( \frac{\pi }{4}-
\frac{\pi k}{2h}\right).
\label{TKk} 
\end{equation}

The coefficients $d_{j}$ as well as the vector 
$\vartheta $ (or numbers $E_{j}$)
fix completely the solution $\phi (y)$ of Eqs.(\ref{EQT},\ref{YN}). They 
determine 
the contribution of the zero modes of the linearized Eq.(\ref{EQT}) and make 
it possible to develop in a standard way the regular expansion 
at large distances. If our scattering theory is consistent with conformal
perturbation theory this expansion should converge to the boundary values 
$\vartheta$ at $y=0$

At the end of this section we note that the analysis of the boundary
Thermodynamic Bethe Ansatz equations with kernels depending on the
reflection coefficients written above gives an exact agreement with
Eq.(\ref {eQ}) for the boundary ground state energy.

\section{Boundary Solutions} \label{sec:boundary-solutions}

It is natural to assume that boundary vacuum solutions can be expressed 
in terms of $\tau$-functions associated with multisoliton solutions of 
ATTs equations. This means that field $\phi$ can be written as:
   
\begin{equation}
\phi (y)=-\sum_{i=0}^{r}e_{i}\log \tau _{i}(y);\quad 
\tau _{i}(y)\rightarrow 0,\ y\rightarrow \infty \quad  \label{t}
\end{equation}
where functions $\tau _{i}(y)$ satisfy the equations:

\begin{equation}
-\tau _{i}^{\prime \prime }(y)+(\tau _{i}^{\prime })^{2}=\left(
\prod_{j=0}^{r}\tau _{j}{}^{\mathcal{I}_{ij}}-\tau _{i}^{2}\right) n_{i}
\label{teq}
\end{equation}
here $\mathcal{I}_{ij}$ is the incidence matrix of the 
extended Dynkin diagram of $G$.

The classical boundary state energy can be expressed in terms of  numbers 
$E_{i}$ and the boundary values of $\tau $-functions \cite{BOWC} as:

\begin{equation}
\mathcal{E}_{bound}^{(cl)}=\frac{h}{2\pi b^{2}n_{i}}\left( \sqrt{n_{i}}E_{i}+
\frac{\tau _{i}^{\prime }(0)}{\tau _{i}(0)}\right) ;\quad i=0,...r
\label{BE}
\end{equation}
Functions $\tau _{i}$ corresponding to multisoliton solutions to the ATT
equations are given by finite order polynomials in the variables

\begin{equation}
Z_{j}=\exp (-m_{j}y)=\exp (-m\nu _{j}y).  \label{Zk}
\end{equation}
The general ansatz for these functions has a form:

\begin{equation}
\tau_{i}(y)=\sum_{k_{1}=0}^{n_{i}}...
\sum_{k_{r}=0}^{n_{i}}Y_{k_{1}...k_{r}}^{(i)}(t_{1}Z_{1})^{k_{1}}...
(t_{r}Z_{r})^{k_{r}}
\label{Yj}
\end{equation}
where $Y_{0...0}^{(j)}=1$ and all other coefficients can be derived from
Eq.(\ref{teq}) if we fix the normalization for the linear in variables 
$Z_{j} $ terms in the expansion (\ref{Yj}): 
\begin{equation}
\tau _{i}(y)=1+\sum_{j=1}^{r}Y_{0..1_{j}..0}^{(i)}t_{j}Z_{j}+O(Z^{2}).
\label{tas}
\end{equation}
These coefficients should satisfy the condition:

\begin{equation}
-\sum_{i=0}^{r}e_{i}Y_{0..1_{j}..0}^{(i)}=N_{j}\xi _{j}.  \label{NC}
\end{equation}
After we fix the factors $N_{j}$ all coefficients $Y^{(i)}$ in Eq.(\ref{Yj})
are universal numbers depending only on $G$. The parameters $t_{j}$ in the
Eq. (\ref{Yj}) are simply related with parameters $d_{j}$ defined by the 
expansion (\ref{fld}). It is easy to see from 
Eqs.(\ref{t},\ref{tas},\ref{NC}) and Eq.(\ref{fld}) that

\begin{equation}
t_{j}=d_{j}/N_{j}  \label{td}
\end{equation}
The parameters $d_{j}$ can be extracted from the boundary scattering theory
described in the previous section. This gives us the possibility to derive
the explicit solution to Eqs.(\ref{EQT},\ref{YN}).

\subsection{$D_{r}$ boundary solution} \label{sec:d_r-bound-solut}

The boundary values of this solution can be described by the numbers $E_{k}$
(see section 3) which in this case are:

\begin{eqnarray}
E_{0} &=&E_{1}=E_{r-1}=E_{r}=\frac{2\sqrt{2}}{h\sin \left( \pi /h\right) }; 
\nonumber \\
E_{k} &=&E_{r-k}=\frac{\cos \left( \pi k/h\right)
\cos \left( \pi(r- k)/h\right) p_{k}^{2}
 p_{r-k}^{2}}{\cos \left(\pi (2k-1)/2h\right)
 \cos \left(\pi (2r-2k-1)/2h\right)}.  \label{dbv}
\end{eqnarray}
These numbers possess all the symmetries of extended 
Dynkin diagram for $D_{r}$. The same is true for the boundary 
solution $\phi (y)$. It means that
in Eq. (\ref{t}) $\tau _{0}=\tau _{1}$ and $\tau _{k}=\tau _{r-k}$. For the
Lie algebra $D_{r}$ the following relation holds:

\begin{equation}
-\sum_{i=0}^{r}n_{i}\cos \left( j\pi (2i-1)/h\right) e_{i}=2\sqrt{h}\sin
(\pi j/h)\xi _{j}  \label{dnor}
\end{equation}
so that it is natural to choose $N_{j}(D)=2\sqrt{h}\sin (\pi j/h)$. It is
convenient also to introduce the functions:

\begin{equation}
\widetilde{\tau }_{0}=\widetilde{\tau }_{1}=\widetilde{\tau }_{r-1}=
\widetilde{\tau }_{r}=\tau _{0}^{2};\quad \widetilde{\tau }_{k}=\tau _{k};\
k=2,...r-2.  \label{ttl}
\end{equation}

The particles which are invariant with respect to symmetries of Dynkin
diagram are particles $2l;\ 2l\leq r-2$ with masses 
$m_{2l}=m\sqrt{8}\sin (2\pi l/h)$. 
The analysis of the boundary $S$
-matrix (\ref{D}) gives that only amplitudes $R_{2l}(\theta )$ have a pole
at $\theta =i\pi /2$. The parameters $t_{2l}=d_{2l}/N_{2l}(D)$ can be
derived from the Eqs.(\ref{D},\ref{D1}) and have a form:

\begin{equation}
t_{2l}=\frac{\tan \left( \pi /4-\pi l/h\right) }{2\cos ^{2}(\pi l/h)}
\prod_{i=1}^{l}\frac{\tan ^{2}(\pi (2i-1)/2h)}{\tan ^{2}(\pi i/h)}
\label{t2l} 
\end{equation}
Parameters $t_{2l}$ and functions $Z_{2l}=\exp (-ym\sqrt{8}\sin (2\pi l/h))$
are defined for $2l\leq r-2$. To write the solution in the most short form
it is convenient, however, to continue these values to $l\leq r-2$. To
continue the parameters $t_{2l}$ to $l\leq r-2$  we can 
use Eq.(\ref{t2l}). In this
way we obtain:

\begin{equation}
t_{h-2l}=-t_{2l};\quad Z_{h-2l}=Z_{2l}
\label{tsym}
\end{equation}
We introduce also $\omega =\exp (2\pi i/h)$. Then the boundary solution can
be written as:

\begin{equation}
\phi (y)=-\half\sum_{i=0}^{r}n_{i}e_{i}\log \widetilde{\tau }_{i}(y)
\label{drs}
\end{equation}
where:
\begin{eqnarray}
\widetilde{\tau }_{i}(y) &=&\sum_{\sigma _{1}=0}^{1}...\sum_{\sigma
_{r-2}=0}^{1}\prod_{k=1}^{r-2}\omega ^{(2i-1)k\sigma _{k}}\left(
t_{2k}Z_{2k}\right) ^{\sigma _{k}}  \nonumber \\
&&\times \prod_{m<n}^{r-2}\left( \frac{\sin \left( \pi (m-n)/h\right) }{\sin
\left( \pi (m+n)/h\right) }\right) ^{2\sigma _{m}\sigma _{n}}  \label{dtau}
\end{eqnarray}
As an example of application of the last equation we give here  the result
for $D_{7}$-solution. This solution is characterized by three function 
$\widetilde{\tau }_{0},\widetilde{\tau }_{2},\widetilde{\tau }_{3}$ 
(\ref{ttl}) which are:
\[
\widetilde{\tau }_{0,3}=
\left( 1 \pm\frac{\sqrt{3}}{2}t_{1}Z_{1}+\frac{1}{2}
t_{2}Z_{2}\pm\frac{\sqrt{3}}{4}\tan ^{2}\left( \frac{\pi }{12}\right)
t_{1}Z_{1}t_{2}Z_{2}\right) ^{2}
\]
\[
\widetilde{\tau }_{2}=1-2t_{2}Z_{2}+\frac{3(t_{1}Z_{1})^{2}}{4}+
\frac{(t_{2}Z_{2})^{2}}{4}+\frac{3}{16}\tan ^{4}
\left( \frac{\pi }{12}\right)
(t_{1}Z_{1}t_{2}Z_{2})^{2}
\]
We note that in the usual basis of roots of $D_{r}$: $e_{0}\phi =-\phi
_{1}-\phi _{2};e_{k}\phi =\phi _{k}-\phi _{k+1},k=1...r-1;e_{r}\phi =\phi
_{r-1}+\phi _{r}$ the solution can be written as $\phi _{k}=\log \left( 
\widetilde{\tau }_{k-1/}\widetilde{\tau }_{k}\right) $. It can be checked
that boundary value of $\phi (y)$ defined by Eqs.(\ref{drs},\ref{dtau})
coincides with $\vartheta $, defined by Eq.(\ref{dbv}) and the classical
boundary ground state energy (\ref{BE}) can be derived using Eq.(\ref{eCl}).

\subsection{$E_{6}$ boundary solution} \label{sec:e_6-bound-solut}

The boundary values for this solution are:
\begin{eqnarray}
E_{0}&=&E_{1}=E_{\overline{1}}=\left( \frac{3+\sqrt{3}}{6}\right)
^{1/2};\quad E_{4}=\frac{\sqrt{3}+1}{6}\left( 3+5/\sqrt{3}\right) ^{1/2}; 
\nonumber \\
E_{2} &=&E_{3}=E_{\overline{3}}=\left( E_{4}E_{0}\right) ^{-1/2}  \label{E6b}
\end{eqnarray}
Here and later we shall use the numeration of nodes $j$ of Dynkin
diagram corresponding to the particles $m_{j}$ (see Fig.s [1-3]) as
follows:

\begin{figure}[ht] \label{fig:e6}
\begin{center}
    \begin{picture}(160,40)(0,0)
      \Line(4,0)(36,0)
      \Line(36,0)(68,0)
      \Line(68,0)(100,0)
      \Line(68,0)(68,16)
      \Line(68,16)(68,32)
      \Line(100,0)(132,0)
      %\Line(132,0)(164,0)
      
      \BCirc(4,0){3}
      \BCirc(36,0){3}
      \BCirc(68,0){3}
      \BCirc(68,16){3}
      \BCirc(68,32){3}
      \BCirc(100,0){3}
      \BCirc(132,0){3}
      
      \Text(180,0)[c]{(Fig.1)}
      \Text(132,-10)[c]{$m_{\overline{1}}$}
      \Text(100,-10)[c]{$m_{\overline{3}}$}
      \Text(68,-10)[c]{$m_4$}
      \Text(36,-10)[c]{$m_3$}
      \Text(4,-10)[c]{$m_1$}
      \Text(78,16)[c]{ $m_2$}
      \Text(78,32)[c]{$0$}
      
    \end{picture}
  \end{center}
\end{figure}

%\vskip 4.0in
\noindent

The solution possesses $\mathbf{Z}_{3}$
symmetry of this diagram. It means that: $\tau _{0}=\tau _{1}=
\tau _{_{\overline{1}}}$ and
 $\tau _{2}=\tau _{3}=\tau _{_{\overline{3}}}$. The
particles that are invariant with respect this symmetry are $m_{2}=m_{ad}$
and $m_{4}$. The analysis of boundary $S$-matrix shows that only amplitude
 $R_{2}=R_{ad}(\theta )$ possesses the pole at $\theta =i\pi /2$ with the
residue (\ref{res}). In this case only one particle contributes to the solution. 
It is
convenient to take $N(E_{6})=2\sqrt{3h}$. Then 
the parameter $t_{2}=d_{ad}/N$ \ is:

\begin{equation}
t_{2}=\tan \left( \frac{\pi }{24}\right) T_{0}T_{2}T_{3}/\sqrt{3}=\tan
\left( \frac{\pi }{24}\right) \tan \left( \frac{5\pi }{24}\right) \tan
\left( \frac{\pi }{12}\right)   \label{t2}
\end{equation}
The variable $Z_{2}=\exp (-m\nu _{2}y)$, where $\nu _{2}^{2}=2(3-\sqrt{3})$
and the solution has a form:
\begin{eqnarray}
\tau _{0} &=&1+(2+\sqrt{3})t_{2}Z_{2};\quad \tau _{2}=1-2t_{2}Z_{2}+\left(
(2+\sqrt{3})t_{2}Z_{2}\right) ^{2};\nonumber \\
\tau _{4} &=&1-3\sqrt{3}t_{2}Z_{2}-3\sqrt{3}(2+\sqrt{3})
(t_{2}Z_{2})^{2}+\left( (2+\sqrt{3})t_{2}Z_{2}\right) ^{3}.
\label{E6S}
\end{eqnarray}
It is easy to check that $\phi (0)=\vartheta $ and the classical boundary 
ground state
energy (\ref{BE}) coincides with (\ref{eCl}).

\subsection{$E_{7}$ boundary solution} \label{sec:e_7-bound-solut}

The boundary values (\ref{Ei}) for this solution are:
\begin{eqnarray}
E_{0} &=&E_{1}=\frac{p_{3}p_{5}}{p_{8}}=\frac{2\sqrt{2}}{9}
\cot \left( \frac{\pi }{9}\right) ;
\ E_{2}=E_{4}=\frac{p_{4}p_{6}}{p_{3}p_{7}};\nonumber \\
E_{3} &=&\frac{p_{5}p_{6}p_{7}}{p_{1}p_{2}p_{3}p_{4}p_{8}};
\ E_{5}=E_{6}=
\frac{p_{1}p_{2}p_{7}}{p_{4}p_{6}};
\ E_{7}=\frac{p_{3}p_{4}p_{8}}{p_{1}p_{2}p_{5}p_{7}}.
\label{E7B}
\end{eqnarray}
Numbers $E_k$ possess $\mathbf{Z}_{2}$ symmetry of the extended  Dynkin 
diagram

\begin{figure}[ht] \label{fig:e7}
\begin{center} 
\begin{picture}(160,30)(0,0)

\Line(-28,0)(36,0)
\Line(4,0)(36,0)
\Line(36,0)(68,0)
\Line(68,0)(100,0)
\Line(68,0)(68,16)
\Line(100,0)(132,0)
\Line(132,0)(164,0)

\BCirc(-28,0){3}
\BCirc(4,0){3}
\BCirc(36,0){3}
\BCirc(68,0){3}
\BCirc(68,16){3}
\BCirc(100,0){3}
\BCirc(132,0){3}
\BCirc(164,0){3}

\Text(200,0)[c]{(Fig.2)}
\Text(164,-10)[c]{ $m_1$}
\Text(132,-10)[c]{ $m_4$}
\Text(100,-10)[c]{ $m_6$}
\Text(68,-10)[c]{ $m_7$}
\Text(36,-10)[c]{ $m_5$}
\Text(68,24)[c]{ $m_3$}
\Text(4,-10)[c]{ $m_2$}
\Text(-28,-10)[c]{ $0$}
\end{picture}  
\end{center}
\end{figure}

 This means that in Eq.(\ref{t}) $\tau_{0}=\tau_{1}$, $\tau_{2}=
\tau_{4}$ and $\tau_{5}=\tau_{6}$ and only $\mathbf{Z}_{2}$ even particles
$m_2=m_{ad}$, $m_4$, $m_6$ and $m_7$ can contribute to this solution.
The analysis of the boundary $S$-matrix (which can be obtained 
by boundary bootstrap fusion procedure from the amplitude $R_f(\theta)$ 
(\ref{dE})) shows that only amplitudes  $R_2=R_{ad}$,  $R_4$ and  $R_7$
possess the poles at $\theta=i\pi/2$.  These amplitudes have a form 
(\ref{gNr}) where the phases $\delta _{j}$ and CDD factors $\Phi _{j}$ are
defined by the functions $\Delta _{j}(E_{7},t)$ and $\psi _{j}(E_{7},t)$ in
Eqs.(\ref{dJ},\ref{cOn},\ref{D1}). For amplitude $R_{2}=R_{ad}$ these
functions are given in section 4. For amplitudes $R_{4}$ and $R_{7}$ they
are:
\begin{eqnarray*}
\Delta _{4} &=&\frac{16\cosh t\cosh 3t\cosh 4t}{\cosh (ht/2)\cosh ht};\ \psi
_{4}=\frac{16\sinh 5t\cosh 4t}{\cosh ht}; \\
\Delta _{7} &=&\frac{8\cosh t\cosh 2t\sinh 6t}
{\sinh t\cosh (ht/2)\cosh ht};\ \psi _{7}=
\frac{8\sinh 3t\sinh 5t\sinh 8t}{\sinh t\sinh 2t\cosh ht}.
\end{eqnarray*}
The coefficients $d_{j}$ can be extracted from these amplitudes using 
Eq.(\ref{dj}). It is convenient to take $N(E_{7})=2\sqrt{3h}$. Then 
the parameters $t_{j}=d_{j}/N$ can be written in terms of numbers $T_k$ 
(\ref{TKk}) as:
\begin{eqnarray}
\sqrt{3}t_{2} &=&\tan \left( \frac{\pi }{2h}\right) 
T_{0}T_{3}T_{5};\ \sqrt{3}t_{4}= 
\tan \left( \frac{\pi }{2h}\right) T_{0}T_{1}T_{3}T_{5}T_{7};
\nonumber \\
\sqrt{3}t_{7} &=&\tan \left( \frac{\pi }{2h}\right)
T_{0}T_{1}^{2}T_{2}T_{3}^{2}T_{5}^{2}T_{7}^{2}(T_{6})^{-1}.
\label{tE7}
\end{eqnarray}
The corresponding variables $Z_j=\exp(-m\nu_j y)$, where:
\begin{eqnarray}
\nu _{2} &=&\sqrt{32}\sin \left( \pi/9\right) 
\sin \left( 2 \pi/9\right) ;\ \nu _{4}=
\sqrt{32}\sin \left(\pi/9 \right) \sin
\left( 4\pi/9 \right) ; \nonumber \\
\ \nu _{7} &=&\sqrt{32}\sin \left(2\pi/9 \right) 
\sin \left( 4\pi/9 \right).
\label{nuE7}
\end{eqnarray}
 
We give here the explicit form for function $\tau_0$. All other 
$\tau$-functions can be derived from it using Eq.(\ref{teq}). For 
example it holds:
\begin{equation}
\tau _{2}=\tau _{0}^{2}-\tau _{0}\tau _{0}^{\prime \prime }+
(\tau_{0}^{\prime })^{2}
\label{tautwo}
\end{equation}
with similar relations for higher $\tau$-functions. 
As an example of application of this equation we give the 
explicit expression for $\tau_2$ in the form (\ref{Yj}) in Appendix A.

For all simply-laced algebras the number $n_0$ is equal to one and 
function $\tau_0$ in Eq.(\ref{Yj}) can be written as:
\begin{equation}
\tau _{0}=\sum_{\sigma _{1}=0}^{1}\sum_{\sigma _{2}=0}^{1}\sum_{\sigma
_{3}=0}^{1}Y_{\sigma _{1}\sigma _{2}\sigma _{3}}^{(0)}(t_{2}Z_{2})^{\sigma
_{1}}(t_{4}Z_{4})^{\sigma _{2}}(t_{7}Z_{7})^{\sigma _{3}}
\label{tau07}
\end{equation}
It is convenient to simplify the notations. The variables 
$\sigma _{i}$ take the values $0$ or $1$. We can  introduce the variable 
$1\leq i$ $\leq 3$ indicating $\sigma _{i}$ which are equal to one. 
The corresponding coefficients $Y_{\sigma_1,.,\sigma_3}^{(0)}$ 
we denote as $y_{i},y_{ij},y_{123}$. 
For example $y_1=Y^{(0)}_{100}$, $y_{13}=Y^{(0)}_{101}$ and so on.
With this notation we have:
\begin{eqnarray}
y_{1} &=&\frac{2\cos ^{2}(2\pi /9)}{\cos (4\pi /9)};\ y_{2}=
\frac{2\cos^{2}(\pi /9)}{\cos (2\pi /9)};\ 
\ y_{3}=\frac{2\cos ^{2}(4\pi /9)}{\cos (\pi/9)}; \nonumber  \\
y_{12} &=&\frac{8\cos ^{4}(2\pi /9)}{\cos (\pi /9)};\ y_{13}=
\frac{8\cos^{4}(4\pi /9)}{\cos (2\pi /9)};\ y_{13}=
\left( 2\cos (4\pi /9\right) )^{7} \nonumber \\
y_{123} &=&(2\cos ^{2}(4\pi /9)/\cos (\pi /9))^{4}
\label{E7cf}
\end{eqnarray}
It can be checked  that
$\phi (0)=\vartheta$, where $\vartheta$ is defined by Eq.(\ref{E7B})   
and the classical boundary ground state
energy (\ref{BE}) coincides with that given by Eq.(\ref{eCl}).

\subsection{$E_{8}$ boundary solution} \label{sec:e_8-bound-solut}
The boundary values for this solutions follow from Eq.(\ref{Ei}), where 
$E_0$ is given by (\ref{E0}) and
\begin{eqnarray}
E_{1} &=&\frac{p_{3}p_{6}p_{10}}{p_{1}p_{5}p_{13}};\ E_{2}=
\frac{p_{8}p_{10}p_{12}}{p_{1}p_{3}p_{5}p_{7}p_{14}};\ E_{3}=
\frac{p_{1}p_{4}p_{5}p_{11}}{p_{3}p_{6}p_{12}}; \nonumber \\
E_{4} &=&\frac{p_{2}p_{3}p_{5}p_{12}p_{13}}{p_{6}p_{7}p_{8}p_{14}};\ E_{5}=
\frac{p_{5}p_{6}p_{12}}{p_{1}p_{2}p_{9}p_{11}};\ E_{6}=\frac{p_{5}p_{9}}
{p_{2}p_{4}p_{8}}; \nonumber \\
E_{7} &=&\frac{p_{1}p_{2}p_{7}p_{8}p_{9}p_{13}}
{p_{4}p_{5}p_{10}p_{10}p_{11}};
\ E_{8}=\frac{p_{4}p_{10}p_{11}p_{14}}{p_{5}p_{9}p_{12}p_{13}}.
\label{E8b}
\end{eqnarray}
\noindent
The extended Dynkin diagram 
\begin{figure}[ht] \label{fig:e8}
\begin{center} 
%\begin{picture}(120,40)(0,80)
\begin{picture}(150,0)(100,24)

\Line(4,0)(36,0)
\Line(36,0)(68,0)
\Line(68,0)(100,0)
\Line(68,0)(68,16)
\Line(100,0)(132,0)
\Line(132,0)(164,0)
\Line(164,0)(196,0)
\Line(196,0)(228,0)

\BCirc(4,0){3}
\BCirc(36,0){3}
\BCirc(68,0){3}
\BCirc(68,16){3}
\BCirc(100,0){3}
\BCirc(132,0){3}
\BCirc(164,0){3}
\BCirc(196,0){3}
\BCirc(228,0){3}

\Text(290,0)[c]{(Fig.3)}
\Text(228,-10)[c]{$0$}
\Text(196,-10)[c]{$m_1$}
\Text(164,-10)[c]{$m_3$}
\Text(132,-10)[c]{$m_5$}
\Text(100,-10)[c]{$m_7$}
\Text(68,-10)[c]{$m_8$}
\Text(36,-10)[c]{$m_6$}
\Text(68,24)[c]{$m_4$}
\Text(4,-10)[c]{$m_2$}

\end{picture}
\end{center} 

\end{figure}
\vskip .250in
\noindent 
in this case has no symmetries and all particles can contribute to the
solution; however the analysis of the boundary $S$-matrix shows that
only amplitudes $R_1=R_{ad}$, $R_2$, $R_5$ and $R_8$ have the poles at
$\theta=i\pi/2$.  The amplitude $R_1=R_{ad}$ is given in section
4. Three other amplitudes with these poles are defined by the
functions $\Delta _{j}(E_{8},t)$ and $\psi _{j}(E_{8},t)$, which have
a form:
\begin{eqnarray*}
\Delta _{2} &=&\frac{16\cosh 3t\cosh 5t\cosh 6t}{\cosh (ht/2)\cosh ht};\
\psi _{2}=\frac{16\sinh 7t\cosh 6t}{\cosh ht}; \\
\Delta _{5} &=&\frac{8\cosh 4t\cosh 5t\sinh 6t}
{\sinh t\cosh (ht/2)\cosh ht};\ \psi _{5}=\frac{8\sinh 3t}
{\cosh ht}\left( \frac{\sinh 11t\sinh 12t}
{\sinh t\sinh 4t}-1\right) ; \\
\Delta _{8} &=&\frac{2\sinh 6t\sinh 10t}{\sinh ^{2}t\cosh (ht/2)\cosh ht};\
\psi _{8}=\frac{8\sinh 5t\sinh 9t\sinh 14t}{\sin t\sinh 2t\cosh ht}.
\end{eqnarray*}
These amplitudes determine the coefficients $d_{j}$  
(see Eq.(\ref{dj})). It is convenient to take $N(E_{8})=2\sqrt{h}$. Then 
the parameters $t_{j}=d_{j}/N$ can be written in terms of numbers $T_k$ 
(\ref{TKk}) as: 
\begin{eqnarray}
t_{1} &=&\tan \left( \frac{\pi }{2h}\right) T_{0}T_{5}T_{9};\ t_{2}=
\tan\left( \frac{\pi }{2h}\right) 
T_{0}T_{3}T_{5}T_{6}T_{9}T_{11}(T_{8})^{-1} \nonumber \\
t_{5} &=&\tan \left( \frac{\pi }{2h}\right) \tan \left( \frac{3\pi }{2h}
\right) T_{7}T_{13}(T_{4}T_{6}T_{8}T_{10}T_{12})^{-1} \nonumber \\
t_{8} &=&\tan ^{2}\left( \frac{\pi }{2h}\right)
T_{1}T_{3}T_{5}T_{7}T_{9}T_{11}T_{13}(T_{6}T_{8}T_{10}^{2}T_{12}^{2})^{-1}.
\label{TE8}
\end{eqnarray}
The corresponding variables $Z_j=\exp(-m\nu_j y)$, where:
\begin{eqnarray}
\nu _{1,5}^{2} &=&3(5-\sqrt{5})/2\mp \sqrt{3(5+\sqrt{5})/2}; \nonumber \\
\nu _{2,8}^{2} &=&3(5+\sqrt{5})/2\mp \sqrt{3(25+11\sqrt{5})/2}.
\label{nuE8}
\end{eqnarray}

We give here the explicit form for function $\tau_0$. All other 
$\tau$-functions can be derived from it using Eq.(\ref{teq}). For 
example function $\tau_1$ is related with $\tau_0$ by Eq.(\ref{tautwo}).
Function $\tau_0$ can be written in the form:
\begin{equation}
\tau _{0}(y)=\sum_{\sigma _{1}=0}^{1}..\sum_{\sigma _{4}=0}^{1}
Y_{\sigma_{1}..\sigma _{4}}^{(0)}(t_{1}Z_{1})^{\sigma _{1}}
(t_{2}Z_{2})^{\sigma
_{2}}(t_{5}Z_{5})^{\sigma _{3}}(t_{8}Z_{8})^{\sigma _{4}}
\label{tttt}
\end{equation}
The coefficients $Y_{\sigma_{1}..\sigma _{4}}^{(0)}$ for this function
are given in Appendix B. It was checked with the accuracy allowed by
{\sf Mathematica}\footnote{Copyright \copyright Wolfram Research Inc.} that
$\phi (0)=\vartheta $ and the classical boundary state energy
(\ref{BE}) coincide with that given by Eq.(\ref{eCl}).

\section{Concluding Remarks} \label{sec:concluding-remarks}

\textit{1}. In the previous sections we studied the semiclassical
asymptotics of integrable boundary ATTs. In particular, we derived the
solution $\phi (y)$ describing the limit $b\rightarrow 0$ of one-point
function $\mathcal{F}_{b}(y)$ of the bulk operator with a boundary.
Here we shortly consider the opposite (dual) limit $b\rightarrow
\infty $ of this function. The boundary value $\Theta $ of function
$\mathcal{F}_{b}(y)$ is given by Eq.(\ref{TET}), and at this limit is
equal

\begin{equation}
\theta ^{\vee }=-\frac{\pi }{2h}\sum_{\alpha >0}\alpha \cot \left( \frac{\pi
\rho _{\alpha }}{h}\right) .  \label{COT}
\end{equation}
For example for the Lie algebra $D_{r}$ this equation can be written as:

\begin{eqnarray}
e_{0}\cdot \theta ^{\vee } &=&e_{1}\cdot \theta ^{\vee }=e_{r-1}\cdot \theta
^{\vee }=e_{r}\cdot \theta ^{\vee }=-\pi/(2h) 
\cot \left( 2\pi /h\right) ;  \nonumber
\\
e_{k}\cdot \theta ^{\vee } &=&-\pi/(2h)
[\cot \left( 2\pi k/h\right) +\cot \left( 2\pi
(r-k)/h\right)].  \label{dbvv}
\end{eqnarray}

At the limit $b\rightarrow \infty $ our theory is described by the weakly
coupled dual theory, which is boundary ATT with Neumann boundary conditions
and function $\phi ^{\vee }(y)=\mathcal{F}_{\infty }(y)$ can be calculated
using the perturbation theory. It can be written in the form:

\begin{equation}
\phi ^{\vee }(y)=\sum_{j=1}^{r}\xi _{j}\left[ d_{j}^{\vee
}e^{-m_{j}y}-\sum_{i=1}^{r}\int \frac{d\theta }{4\sqrt{h}}
\frac{f_{ii}^{j}
\sin (\theta _{ii}^{j})e^{-2m_{i}y\cosh \theta }}{\cosh ^{2}\theta
-\cos ^{2}(\theta _{ii}^{j}/2)}\right]   \label{fid}
\end{equation}
Here $f_{ii}^{j}$ are the fusion constants and $\theta _{ii}^{j}$ are the
fusion angles. The constants $f_{ii}^{j}$ are different from zero and equal
to $1$ or $-1$ only if particle $j$ is a bound state of two particles $i$.
In this case the fusion angle is defined by relation: $m_{j}=2m_{i}\cos
(\theta _{ii}^{j}/2)$. The complete list of constants $f_{ii}^{j}$ and
angles $\theta _{ii}^{j}$ can be found in Ref.\cite{CDB}. The constants 
$d_{j}^{\vee }$ describing one particle contributions are different from zero
exactly for the same particles $j$ that and the constants $d_{j}$ calculated
in previous section. For these particles they can be expressed in terms of
residues of amplitudes $R_{j}(\theta )$ (see Eq.(\ref{dj})) as:

\begin{equation}
d_{j}^{\vee }=\lim_{b\rightarrow \infty }\sqrt{\pi }bD_{j}(b)=\frac{\pi }
{\sqrt{h}}  \label{ddual}
\end{equation}

As an example of the application of Eq.(\ref{fid}) we give the explicit
expression for function $\phi ^{\vee }(y)$ in $D_{r}$ ATTs:

\begin{equation}
\phi ^{\vee }(y)=\frac{\pi }{\sqrt{h}}
\sum_{l=1}^{[r/2]-1}\xi _{2l}
\left[e^{-m_{2l}y}-\int \frac{d\theta }{4\pi }
(g_{l}-g_{r-1-l}+f_{l})\right]
\label{dddd}
\end{equation}
where

\begin{eqnarray}
g_{l}(\theta ) &=&\sin (2\pi l/h)
\frac{\exp (-4\sqrt{2}\sin (\pi l/h)my
\cosh\theta )}{\cosh ^{2}\theta -\cos ^{2}(\pi l/h)};  \nonumber \\
\ f_{l}(\theta ) &=&2\sin (4\pi l/h)\frac{\exp (-2\sqrt{2}my\cosh \theta )}
{\cosh ^{2}\theta -\sin ^{2}(2\pi l/h)}  \label{gf}
\end{eqnarray}
A plot of the  functions $\phi_0=-e_0\cdot\phi$ and 
$\phi_0^\vee=-e_0\cdot\phi^\vee$ for $D_7$ ATT
is given in Fig.4.

\setcounter{figure}{3}

\begin{figure}[ht] 
  \begin{center}
    \mbox{\epsfig{file=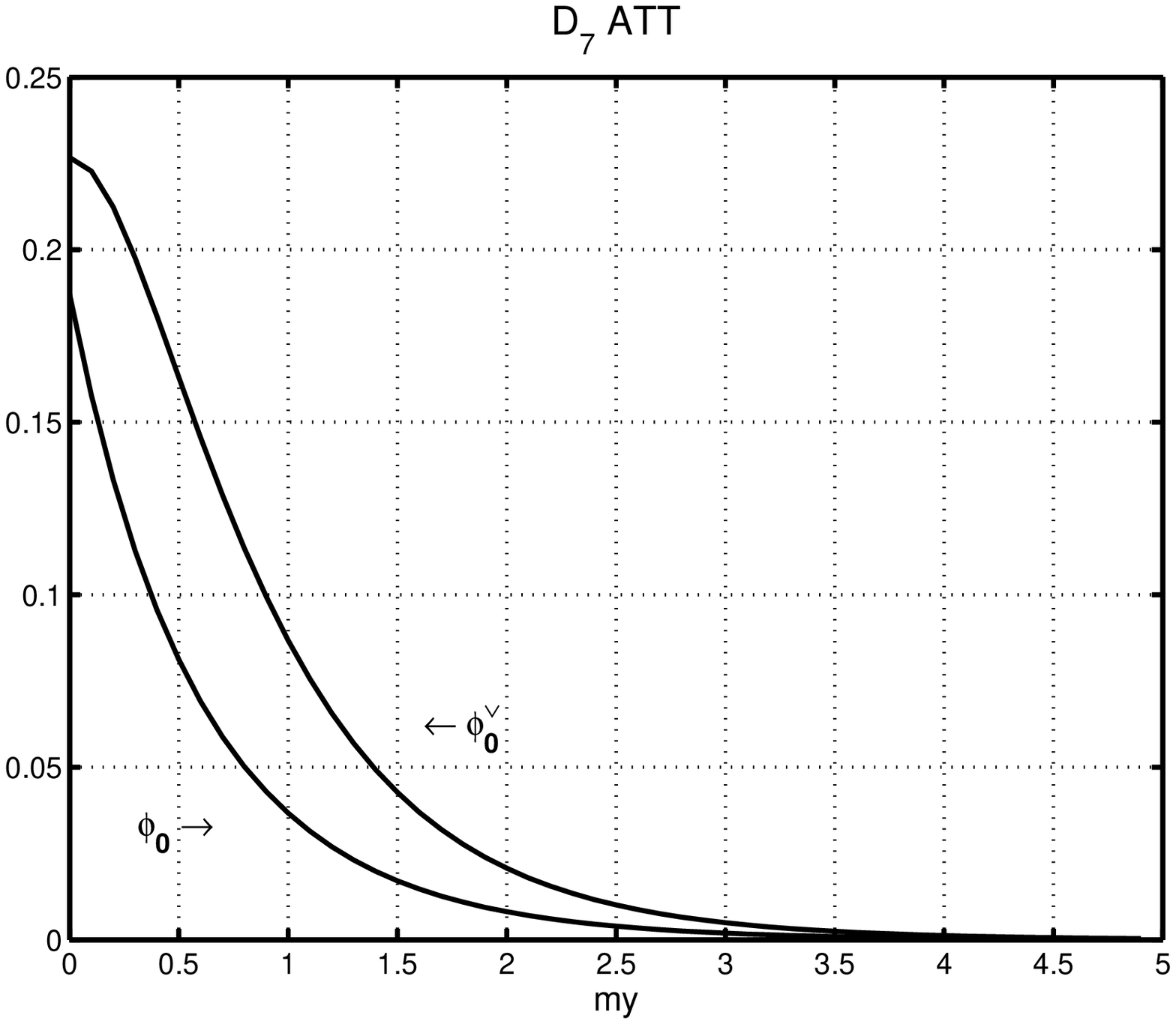,width=10.cm}}
 \caption{\sl The functions $\phi_0=-e_0\cdot\phi$ and
 $\phi^\vee_0=-e_0\cdot\phi^\vee$ for the $D_7$ ATT.}
  \end{center}
\end{figure}
It is straightforward to derive from equations (\ref{dddd},\ref{gf}) that
\begin{equation}
\phi ^{\vee }(0)=\frac{\pi }{2\sqrt{h}}\sum_{l=1}^{[r/2]-1}\xi _{2l}
\left( 1-\frac{4l}{h}\right)  \label{lh}
\end{equation}
If we take into account Eq.(\ref{dnor}) we can easily find that boundary
value $\phi ^{\vee }(0)$ coincides with vector $\theta ^{\vee }$ defined by
Eq.(\ref{dbvv}). The same result can be obtained for all simply laced ATTs.

\textit{2}. In this paper we derived the weak and strong coupling
limits of function $\mathcal{F}_{b}(y)$. In the intermediate region
the long distance behavior of this function can be expressed in terms
of boundary amplitudes $R_{j}$ and form factors \cite{DPWT}. Some of
form factors in ATTs are calculated in Ref.\cite{SL}. At short
distances we can calculate $\mathcal{F}_{b}(0)$ which is equal to
vector $\Theta $. Using boundary conditions and equations of motion we
can express first two derivatives of this function through the
boundary \ one-point function $G_{B}(a):$

\begin{equation}
\mathcal{F}_{b}^{\prime }(0)=\frac{b^{2}\mu _{B}}{2}
\sum_{j=0}^{r}e_{j}G_{B}(be_{j});\ \mathcal{F}_{b}^{\prime \prime
}(0)=b^{2}\mu \sum_{j=0}^{r}e_{j}G_{B}(2be_{j}).  \label{der}
\end{equation}
To calculate the next derivatives of this function we need the boundary VEVs
of the descendant fields. At present the problem of their calculation is not
solved yet. We hope to return to this problem in the next publications.

\textit{3}. In the main part of this paper we studied boundary 
simply laced ATTs with
dual to Neumann boundary conditions (\ref{MND}). At the end of this section
we consider the classical limit of non-simply laced ATTs with the same
boundary conditions. These theories are dual to the ATTs with dual affine
algebras and boundary conditions (\ref{N}). In the non-simply laced case the
corresponding boundary vacuum solutions $\phi (y)$ satisfy the same Eqs.(\ref
{EQT}) with modified boundary conditions at $y=0$ which can be written as:

\begin{equation}
\partial _{y}\phi =m\sum_{i=0}^{r}\sqrt{2n_{i}/e_{i}^{2}}e_{i}\exp
(e_{i}\cdot \phi ).  \label{MBC}
\end{equation}

The solutions $\phi _{nsl}$ of these equations can be obtained from the
solutions $\phi _{sl}$ for simply laced ATTs described in the previous
section by the reduction with respect to the symmetries of extended $DE$
Dynkin diagrams (which are the symmetries of solutions $\phi _{sl}$). This
reduction can be described in the following way. Affine Dynkin diagrams for
non-simply laced algebras can be obtained by folding the nodes of diagrams
for simply laced algebras $G$. Let us the set of nodes $\{j\}$ of affine
Dynkin diagram for $G$ corresponds after the folding to the node $j$ of the
diagram for non-simply laced algebra $\mathcal{G}$. We denote as $e_{J}$ the
root of $G$ with arbitrary $J\in \{j\}$ and as $e_{j}$ the corresponding
root of $\mathcal{G}$. Then it holds:

\begin{equation}
e_{j}\cdot \phi _{nsl}(y)=e_{J}\cdot \phi _{sl}(y)  \label{nsl}
\end{equation}
This relation defines completely the solutions of Eqs.(\ref{EQT},\ref{MBC})
for all for non-simply laced algebras $\mathcal{G}$. The ratio of
corresponding boundary ground state energies is equal to the ratio of
Coxeter numbers of $\mathcal{G}$ and $G$.

Extended Dynkin diagram for the Lie algebra $B_{r}$ can be 
obtained by folding nodes $r$ and 
$r+1$ of the diagram for $D_{r+1}$. In this case $\mathcal{E}^{(cl)}(B_{r})=
\mathcal{E}^{(cl)}(D_{r+1})$. In the dual limit $B_{r}$ ATT with boundary
conditions (\ref{MBC}) is described by weakly coupled $B_{r}^{\vee }$ (or 
$A_{2n-1}^{(2)}$) ATT with Neumann boundary conditions. The perturbative
calculations and analysis of boundary $S$-matrix for this theory gives us
the reasons to conjecture that in quantum case the boundary ground state 
energies for
$B_{r}$ and $D_{r+1}$ ATTs have a similar form:

\begin{equation}
\mathcal{E}_{bound}^{(q)}(B_{r})=\frac{m\cos (\pi /4-\pi /2h)}{4\sin (\pi
x/2h)\cos (\pi (1-x)/2h)}  \label{bdbe}
\end{equation}
where $h$ now depends on $x$: $h=2r-x$.

Folding the nodes $r+1$ with $r+2$ and $0$ with $1$ of $D_{r+2}$ diagram we
obtain the diagram for $C_{r}^{\vee }$ or ( $D_{r+1}^{(2)}$) ATT. The
similar consideration of this theory leads us to conjecture that 
$\mathcal{E}_{bound}^{(q)}(C_{r}^{\vee })$ is also described by 
Eq.(\ref{bdbe}) where
now $h=2r+2(1-x)$.

Folding the nodes $k$ and $2r-k$ of affine $D_{2r}$ diagram we obtain the
classical solution for $B_{r}^{\vee }$ boundary ATT. In this case 
$\mathcal{E}^{(cl)}(B_{r}^{\vee })=\mathcal{E}^{(cl)}(D_{2r})/2$. 
The solution for 
$BC_{r}$ (or $A_{2n}^{(2)}$) ATT can be obtained from $D_{2r+2}$ solution by 
folding $D_{2r+2}$ diagram with respect to all symmetries. In this case all 
$r$ particles contribute to the solution and $\mathcal{E}^{(cl)}(BC_{r})=
\mathcal{E}^{(cl)}(D_{2r+2})/2$.

The solution for $G_{2}$ ATT can be obtained from $D_{4}$ solution by
folding the nodes $1,3,4$ of affine $D_{4}$ Dynkin diagram. In this case
only the heavy particle $m_{2}$ contributes to the 
solution and $\mathcal{E}^{(cl)}(G_{2})=\mathcal{E}^{(cl)}(D_{4})$. 
Contrary to that only the light
particle $m_{1}$ contributes to the solution for $G_{2}^{\vee }$ or 
($D_{4}^{(3)}$) ATT. This solution can be obtained by the reduction with
respect threefold symmetry of $E_{6}$ extended Dynkin diagram. In this case 
$\mathcal{E}^{(cl)}(G_{2}^{\vee })=\mathcal{E}^{(cl)}(E_{6})/3$. Folding the
nodes $1$ with $\overline{1}$ and $3$ with $\overline{3}$ of the same
diagram we obtain ``one particle'' solution for $F_{4}$ ATT. The ``three
particle'' solution for $F_{4}^{\vee }$ or ($E_{6}^{(2)}$) ATT can be
derived by the reduction with respect to $\mathbf{Z}_{2}$ symmetry of $E_{7}$
extended Dynkin diagram. As a result $\mathcal{E}^{(cl)}(F_{4}^{\vee })=
\mathcal{E}^{(cl)}(E_{7})/2$. The solution for the last ATT corresponding to 
the Lie algebra $C_{r}$ vanishes: $\phi (y)=0$.

In this paper we found explicitly all solutions of
Eqs.(\ref{EQT},\ref{MBC}).  However some of non-simply laced ATTs 
require more general integrable boundary conditions \cite{BCD}, which
form the parametric families. We intend to describe the corresponding
boundary one-point functions, scattering theory and boundary solutions
in a separate publication.

\bigskip

\begin{center}
\textbf{Acknowledgments}
\end{center}

We are grateful to Al. Zamolodchikov V.Riva and B.Ponsot for useful
discussions.  E.O. warmly thanks A. Neveu, director of L.P.M.,
University of Montpellier II, and all his colleagues for the kind
hospitality at the Laboratory.  This work supported by part by the EU
under contract ERBFRMX CT 960012 and grant INTAS-OPEN-00-00055.

\begin{center}
\bigskip \textbf{Appendix A} \label{APP_A}
\end{center}

In this appendix we give the explicit form for the function  $\tau _{2}$ in 
$E_{7}$ ATT. This function can be obtained from the function $\tau _{0}$ 
(see Eqs.(\ref{tau07},\ref{E7cf})
using Eq. (\ref{tautwo}) and is determined by the following coefficients 
$Y_{abc}^{(2)},a,b,c=0,1,2$; $Y_{000}^{(2)}=1$:

\begin{equation}
\tau_{2}=
\sum\limits_{a=0}^{2}\sum\limits_{b=0}^{2}\sum\limits_{c=0}^{2}Y_{abc}^{(2)}
(t_{2}Z_{2})^{a}(t_{4}Z_{4})^{b}(t_{7}Z_{7})^{c}
\label{taus}
\end{equation}
where

$Y_{100}^{(2)}=y_{1}^{(2)}=4\cos (2\pi /9);\
Y_{010}^{(2)}=y_{2}^{(2)}=-4\cos (\pi /9);$

$Y_{001}^{(2)}=y_{3}^{(2)}=-4\cos (4\pi /9);\quad Y_{110}^{(2)}=
1/\cos^{2}(\pi /9);\ Y_{101}^{(2)}=-1/\cos ^{2}(2\pi /9);$

$Y_{011}^{(2)}=-4\cos ^{2}(4\pi /9)/\cos ^{2}(\pi /9);\
Y_{111}^{(2)}=128\cos ^{4}(4\pi /9)/\cos ^{2}(\pi /9); $

$Y_{200}^{(2)}=y_{1}^{2};\ Y_{020}^{(2)}=y_{2}^{2};\
Y_{002}^{(2)}=y_{3}^{2};\quad Y_{210}^{(2)}=y_{12}y_{2}^{(2)}y_{1}/y_{2};$

$Y_{120}^{(2)}=y_{12}y_{1}^{(2)}y_{2}/y_{1};\
Y_{201}^{(2)}=y_{13}y_{3}^{(2)}y_{1}/y_{3};\
Y_{102}^{(2)}=y_{13}y_{1}^{(2)}y_{3}/y_{1};$

$Y_{021}^{(2)}=y_{23}y_{3}^{(2)}y_{2}/y_{3};\
Y_{012}^{(2)}=y_{23}y_{2}^{(2)}y_{3}/y_{2};\quad Y_{220}^{(2)}=y_{12}^{2};\
Y_{202}^{(2)}=y_{13}^{2};$

$Y_{022}^{(2)}=y_{23}^{2};\quad
Y_{211}^{(2)}=y_{12}y_{13}Y_{011}^{(2)}/(y_{2}y_{3});\
Y_{121}^{(2)}=y_{12}y_{23}Y_{101}^{(2)}/(y_{1}y_{3});$

$Y_{112}^{(2)}=y_{13}y_{23}Y_{110}^{(2)}/(y_{1}y_{2});\quad
Y_{221}^{(2)}=y_{123}y_{12}y_{3}^{(2)}/y_{3};\quad
Y_{212}^{(2)}=y_{123}y_{13}y_{2}^{(2)}/y_{2};$

$Y_{122}^{(2)}=y_{123}y_{23}y_{1}^{(2)}/y_{1};\quad
Y_{222}^{(2)}=y_{123}^{2}$.

\newpage

\begin{center}
\textbf{Appendix B} \label{APP_B}
\end{center}

In this appendix we list the coefficients $Y_{\sigma _{1}\sigma _{2}\sigma
_{3}\sigma _{4}}^{(0)}$ in the Eq.(\ref{tttt}) for $\tau _{0}$ function in 
$E_{8}$ ATT. It is convenient to simplify the notations. The variables 
$\sigma _{i}$ take the values $0$ or $1$. We can introduce the variable 
$1\leq i$ $\leq 4$ indicating $\sigma _{i}$ which are equal to one.
The corresponding coefficients 
$Y_{\sigma _{1}..\sigma _{4}}^{(0)}$ We denote as $y_{i}$, $y_{ij}$, 
$y_{ijk}$ and $y_{1234}$. For example, 
$y_{2}=Y_{0100}^{(0)};$ $y_{13}=Y_{1010}^{(0)}$ and so on. Then 
$Y_{0000}^{(0)}=1$ and

$y_{1}=\frac{(1+\sqrt{5})^{2}}{8}\left( \sqrt{3}+\sqrt{\frac{5}{3}}+
\sqrt{4+\frac{8}{\sqrt{5}}}\right);
\ y_{2}=\frac{(1+\sqrt{5})}{4}
\left( \frac{7}{\sqrt{3}}-
\sqrt{\frac{5}{3}}+\sqrt{4+\frac{8}{\sqrt{5}}}\right) ;$

$y_{3}=\frac{(1+\sqrt{5})^{2}}{8}\left( \sqrt{3}+\sqrt{\frac{5}{3}}-
\sqrt{4+\frac{8}{\sqrt{5}}}\right) ;\ y_{4}=\frac{(1+\sqrt{5})}{4}
\left( \sqrt{\frac{5}{3}}-\frac{7}{\sqrt{3}}+
\sqrt{4+\frac{8}{\sqrt{5}}}\right) ;$

$y_{12}=\frac{43}{3}-\frac{19}{\sqrt{5}}+\sqrt{\frac{4289}{3}-
\frac{9362}{3\sqrt{5}}};\ y_{13}=-\frac{44}{3}+\frac{83}{\sqrt{5}}-
4\sqrt{\frac{298}{3}-\frac{458}{3\sqrt{5}}};$

$y_{14}=\frac{71}{3}-\frac{101}{\sqrt{5}}+\sqrt{\frac{7801}{3}-
\frac{14342}{3\sqrt{5}}};\ y_{23}=\frac{25}{3}-
\frac{19}{\sqrt{5}}+5\sqrt{\frac{17}{3}-\frac{38}{3\sqrt{5}}};$

$y_{24}=-\frac{52}{3}+\frac{43}{\sqrt{5}}-4\sqrt{\frac{125}{3}-
\frac{278}{3\sqrt{5}}};\ y_{34}=-\frac{43}{3}+
\frac{19}{\sqrt{5}}+\sqrt{\frac{4289}{3}-
\frac{9362}{3\sqrt{5}}};$

$y_{123}=y_{12}y_{13}y_{23}/(y_{1}y_{2}y_{3});\
y_{124}=y_{12}y_{14}y_{24}/(y_{1}y_{2}y_{4});\quad $

$y_{134}=y_{13}y_{14}y_{34}/(y_{1}y_{3}y_{4});$\ 
$y_{234}=y_{23}y_{24}y_{34}/(y_{2}y_{3}y_{4});\quad $

$y_{1234}=y_{12}y_{13}y_{23}y_{14}y_{24}y_{34}/(y_{1}y_{2}y_{3}y_{4})^{2}$.

$\quad \quad $

\bigskip

\begin{center}
\textbf{Appendix C}  \label{APP_C}
\end{center}

The classical vacuum solutions of Sec.\ref{sec:boundary-solutions} have
been cross-checked numerically. We used {\sf
Mathematica}\footnote{Copyright \copyright Wolfram Research Inc.}'s
routine {\tt NDSolve} which allow good accuracy control through the
parameters {\tt Prec, AccuracyGoal, WorkingPrecision}. 
\begin{figure}[ht] 
  \begin{center}
    \mbox{\epsfig{file=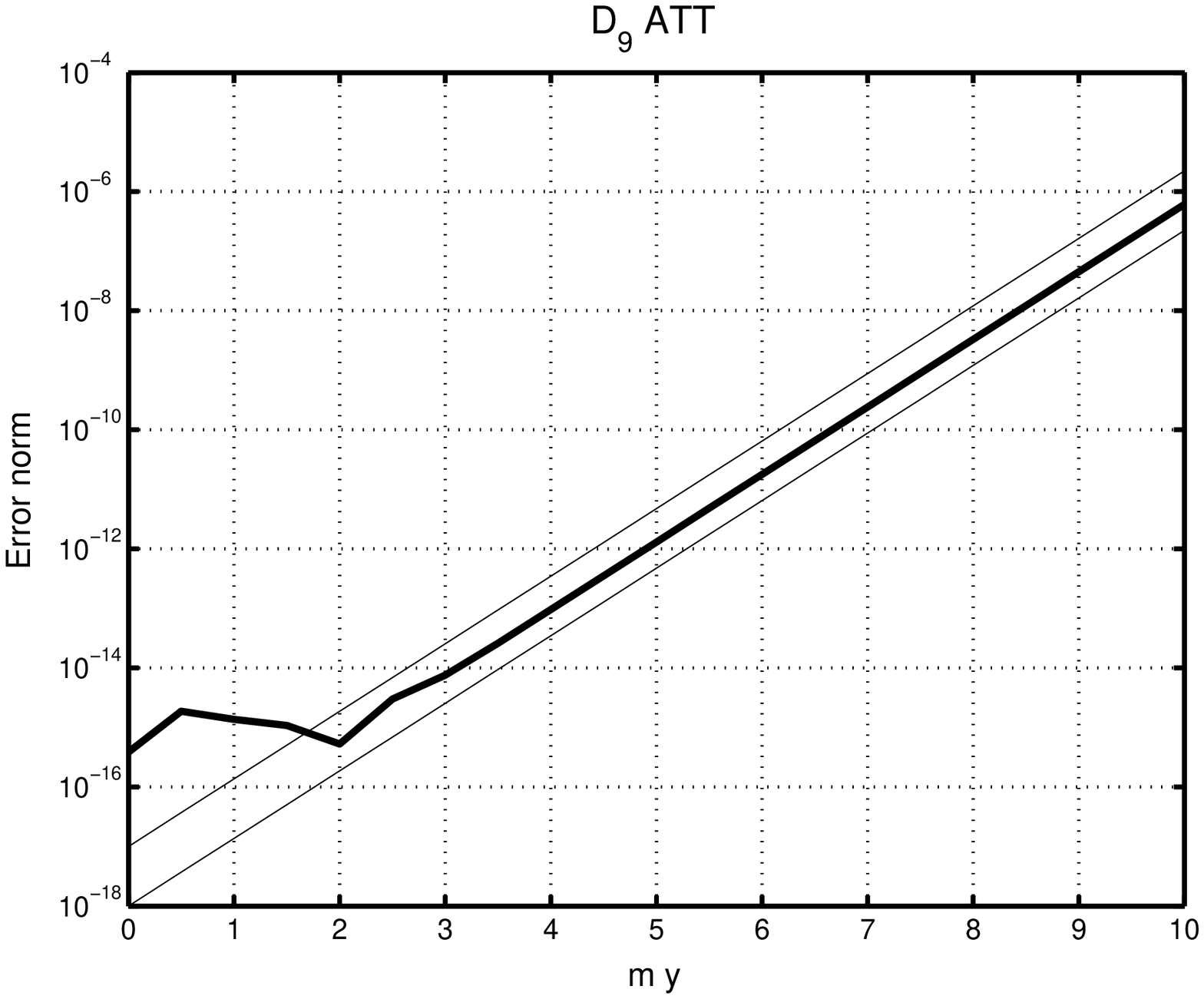,width=10.cm}}
 \caption{\sl\small The exponential growth of the difference between
 exact and numerical solution\/: the slope is given by the square root
of the highest eigenvalue of the mass matrix.}
  \end{center}
\end{figure}
The code for the 
solution of boundary problem ({\ref{EQT},\ref{YN}) has
been developed for all cases considered in
this paper; it is available at {\tt
http://www.fis.unipr.it/$\sim$onofri}.

 The inherent instability of the boundary value problem (i.e. the
existence of solutions with exponential growth at infinity) makes the
direct numerical problem rather ill-conditioned; however this fact is
completely under control. The numerical solutions agree with the exact
ones until the unstable modes eventually take over by amplifying the
unavoidable truncation and algorithmic errors. In Fig.5 we may see
that the discrepancy between exact and numerical solution eventually
grows with exactly the slope dictated by the square root of the
largest eigenvalue of the mass matrix (the pattern is replicated
systematically for all the other Toda systems).

In case an exact solution were not available, a possible strategy,
is given by ``backward-integration'': starting from large
values of $y$ and assuming a general linear combination $\phi=\sum
d_j\xi_{j}\exp\{-m\nu_j\,y\}$ one can integrate back to $y=0$ and fit the
coefficients $d_j$ to match the boundary conditions. This technique is
not affected by the unstable modes and can give rather accurate values
for the asymptotic data.

\end{document}